\documentstyle[sprocl,epsfig]{article}

\def\NPB{{\em Nucl. Phys.} B}
\def\PLB{{\em Phys. Lett.}  B}

\def\ZPC{{\em Z. Phys.} C}

\def\be{\begin{equation}}
\def\ee{\end{equation}}
\def\bea{\begin{eqnarray}}
\def\eea{\end{eqnarray}}
\def\beq{\begin{equation}}
\def\eeq{\end{equation}}
\def\bea{\begin{eqnarray}}
\def\eea{\end{eqnarray}}
\def\bem{\begin{math}}
\def\eem{\end{math}}
\def\bit{\begin{itemize}}
\def\eit{\end{itemize}}
\def\bla{\begin{flushright}}
\def\ela{\end{flushright}}
\def\qq2{$Q^2$}               % Q2
\def\aa1{$A_1(x,Q^2)$}        % A1
\def\ff1{$F_1(x,Q^2)$}        % F1
\def\gg1{$g_1(x,Q^2)$}        % g1

\begin{document}

\title{$F_2$ STRUCTURE FUNCTION AND HIGHER-TWIST CONTRIBUTIONS 
(NONSINGLET CASE)}

\author{A. V. KOTIKOV, V. G. KRIVOKHIJINE}

\address{Particle Physics Laboratory, JINR \\ Dubna, 141980 Russia \\
E-mail: kotikov@sunse.jinr.ru, kvg@sunse2.jinr.ru}


\maketitle\abstracts{ 
We performed the combine fits of SLAC, BCDMS and NMC data at NLO level.
The data are fitted very well with different
%the both: renormalon-type and non-renormalon type
parametrizations of higher twist contributions.
The theorety-based
%renormalon-type
parametrizations lead to the low values of QCD
parameter $\Lambda^{f=4}_{\overline{MS}} \sim 140 \pm 21~ MeV$
(or $\alpha_s(M^2_Z) \sim 0.103 \pm 0.002$).}

%\section{Guidelines}

%\subsection{Producing the Hard Copy}\label{subsec:prod}

%The deep inelastic scattering (DIS) leptons on hadrons is the basical
% process to study the values of the partonic distribution functions
%which are universal (after choosing of a renormalization scheme) and
%may to use in other processes. 
The accuracy of the present data for deep inelastic
(DIS) structure functions (SF) reached the level at which
the $Q^2$-dependence of logarithmic QCD-motivated and power-like ones
%are observed and 
may be studied separately (see, for example, the
recent reviews \cite{Maul} and references therein).
In the present letter we analyse at NLO order of perturbative QCD
the most known DIS SF 
$F_2(x,Q^2)$ taking into account SLAC, BCDMS and NMC
\cite{data}
experimental data. We
stress the power-like effects, so-called twist-4 (i.e.
$\sim 1/Q^2$) and twist-6 (i.e. $\sim 1/Q^4$) contributions.
To our purposes we represent the SF $F_2(x,Q^2)$ as the contribution
of the part $F_2^{pQCD}(x,Q^2)$ (including  target mass 
corrections (TMC))
described by perturbative QCD and the  
nonperturabative part:
\begin{equation}
F_2(x,Q^2) 
%\equiv F_2^{full}(x,Q^2)
=F_2^{pQCD}(x,Q^2)+\frac{h_4(x)}{Q^2} + \frac{h_6(x)}{Q^4}
\label{1}
\end{equation}
%where SF $ F_2^{pQCD}(x,Q^2)$ contains also the effect of target mass 
%corrections (TMC).

%{\bf 1.}~The SF $F_2^{pQCD}(x,Q^2)$ obeys the perturabative QCD dynamics
%including the target mass corrections (and coincides with $F_2^{NS}(x,Q^2)$
%when the target mass corrections are withdrawn).
Contrary to standard fits (see, for example, \cite{VM}) when the 
direct
numerical calculations based on 
%Dokshitzer-Gribov-Lipatov-Altarelli-Parisi
DGLAP equation 
%\cite{DGLAP} 
are used to evaluate of SF, 
we use the exact solution of DGLAP equation
for the Mellin moments of 
%$F_2^{full}(x,Q^2)$ (and analogouslyfor SF  
$F_2^{pQCD}(x,Q^2)$
%and $F_2^{NS}(x,Q^2)$:
%\begin{equation}
%M_n^{pQCD}(Q^2)=\int_0^1 x^{n-2}F_2^{pQCD}(x,Q^2)dx~~~~ 
%(\mbox{ hereafter } k=full, PQCD, NS, ...)
%\label{2}
%\nonumber
%\end{equation}
and
the subsequent reproduction of $F_2(x,Q^2)$ 
%and/or $F_2^{pQCD}(x,Q^2)$ at every needed 
at some $Q^2$-value with help of the Jacobi 
Polynomial expansion method \cite{Kri,SIPT,NNLO}
%(see similar analyses at the next-to-leading (NLO) level \cite{Kri}-\cite{KS}
%and at the next-next-to-leading (NNLO) level \cite{PKK}-\cite{KKPS2}).\\0
%
%%{\bf 2.}~
%The method of the Jacobi polynomial expansion was developed
%in \cite{Barker, Kri} and described in Refs.\cite{Kri}. Here we introduce only
%some basical definitions.
%
%Having the QCD expressions for the Mellin moments
%$M_n^{NS}(Q^2)$ we can reconstruct the SF $F_2(x,Q^2)$
%%using the Jacobi polynomial expansion method:
%as
\begin{equation}
F_{2}^{N_{max}}(x,Q^2)=x^{\alpha}(1-x)^{\beta}\sum_{n=0}^{N_{max}}
\Theta_n ^{\alpha , \beta}
(x)\sum_{j=0}^{n}c_{j}^{(n)}{(\alpha ,\beta )}
M_{j+2} \left ( Q^{2}\right ),
%\label{4.1}
\nonumber
\end{equation}
where $\Theta_n^{\alpha,\beta}$ are the Jacobi polynomials
and $\alpha,\beta$ are their parameters, fixed by the condition
of the requirement of the minimization of the error of the
reconstruction of the
structure functions.\\
% (see Ref.\cite{Kri} for details).

For $n$-space 
%the equation corresponding to 
 eq.(\ref{1}) transforms to
%has the form 
\begin{equation}
M_n(Q^2) 
=M_n^{pQCD}(Q^2)+ \biggl[
\frac{C_4(n)}{Q^2} + \frac{C_6(n)}{Q^4} \biggr] \cdot M_n^{NS}(Q^2),
\label{1.1}
\end{equation}
where (hereafter ($a=4, 6$)) 
$ C_a(n) = \int_0^1 x^{n-2}h_a(x)dx/M_n^{NS}(Q^2) $. 
The $Q^2$-evolution of the moments $M_n^{NS}(Q^2)$ is given by the well
known perturbation QCD formulae (see \cite{Buras}).
The moments $M_n^{pQCD}(Q^2)$ are
distinct from $M_n^{NS}(Q^2)$ by including TMC.
The moment $M_n^{NS}(Q_0^2)$
is theoretical input of our analysis 
and its parameters
should be found together with $h_4(x)$, $h_6(x)$ and QCD
parameter $\Lambda $ by the fits of experimental data.

%{\bf 2.}~
The shapes $h_a(x)$  of the higher twist (HT)
corrections  are of primary consideration in our
analysis. They are choosen in the several different forms.
\begin{itemize}
\item 
The twist-4 and twist-6 terms are fixed in agreement with the infrared 
renormalon model (see \cite{DW} and references therein).
%Then, the
The values of  HT coefficient functions
$C_4(n)$ and $C_6(n)$ are given (see \cite{DW})
%:
in the form $C_a^{ren}(n) = B_a(n) \cdot A_a^{ren}$, where $B_a(n)$ are
known $n$-dependent functions and 
%
%\begin{eqnarray}
%%C_4(n) &=&
%C_4^{ren}(n)
%&=& \biggl(n +4 -\frac{2}{n+1} -\frac{4}{n+2} - 4 S_1(n) \biggr) A_4^{ren}
%%~~~~~~~\mbox{ and } 
%\nonumber \\
%%\label{8} \\
%%C_6(n) &=& 
%C_6^{ren}(n) &=&
%\biggl(\frac{1}{2} n^2 - \frac{3}{2} n +16
% -\frac{4}{n+1} -\frac{36}{n+3} - 4 S_1(n) \biggr) A_6^{ren}
%% \label{9} 
%\nonumber
%\end{eqnarray}
%with
%$S_1(n)=\sum^n_{j=1} 1/j $ and
$A_a^{ren}$ are some coefficients. The values of
$A_a^{ren}$ have been estimated (in \cite{DW}) as $A_4^{ren}=0.2$ GeV$^2$ and  
$A_6^{ren}= (A_4^{ren})^2=0.04$ GeV$^4$. We use them here as free parameters.
\item 
The twist-4 term in the form 
$ h_4(x)\sim \frac{d}{dx}lnF_2^{NS}(x,Q^2) \sim 1/(1-x) $.
This behaviour matches 
%(see \cite{CDR}) 
the fact that higher twist effects are usually important only at higher $x$. 
The twist-4 coefficient function has the form
%\begin{eqnarray}
$
%C_4(n) =
C_4^{der}(n)
= (n -1) A_4^{der}$.
% \label{12} 
%\end{eqnarray}
%
\item
%{\it Case B.} The higher twist terms 
%is fixed in the agreement with the IRR model 
%(see the Case A) and the twist-6 one is represented in the form
HT terms
are considered as free parameters at each $x_i$ bin. They have the form
%\begin{equation}
$h_a^{free}(x)=\sum_{i=1}^{I} h_a(x_i)$, 
%\label{9.1}
%\end{equation}
where $I$ is the number of bins.
The
constants $h_a(x_i)$ (one per $x$-bin) parametrize $x$-dependence of 
$h_a^{free}(x)$.
\end{itemize}
%%%\\
%We put
%\begin{eqnarray}
%x_i=0.03,~0.05,~0.08,~0.15,~0.25,~0.35,~0.45,~0.50,~0.55,~0.65,~0.80 \nonumber
%\\
%\mbox{ for }~~ i=1,2,...,11
%\label{9.2}
%\end{eqnarray}\\

To clear up the importance of HT terms we fit SLAC, BCDMS and NMC 
($H_2$ and $D_2$) data \cite{data} (including the systematical errors),
keeping identical form of perturbative part
at NLO approximation.
We choose hereafter $Q^2_0$ = 10 GeV$^2$, $N_{max} =6$, $0.3 \leq x \leq 0.9$
and $I=6$. Basic characteristics of the quality of the fits are
$\chi^2/DOF$ for SF $F_2$ and for its slope $dlnF_2/dlnQ^2$, which has
very sensitive 
perturbative properties (see \cite{Buras}). We use MINUIT program for
minimization of two $\chi^2 $ values:
\begin{eqnarray}
\chi^2(F_2) = {\biggl|\frac{F_2^{exp} - F_2^{teor}}{\Delta F_2^{exp}}
\biggr| }^2 ~ \mbox{ and }  ~
\chi^2(\mbox{slope}) = {\biggl|\frac{b^{exp} - b^{teor}}{\Delta b^{exp}}
\biggr| }^2,~ \biggl(
b=\frac{dlnF_2}{dlnQ^2} \biggr)
\nonumber
\end{eqnarray}
%where $b=dlnF_2/dlnQ^2$.

We obtain the following results:
\begin{itemize}
\item Only perturbative QCD part is included 
%No TMC and HT corrections
(i.e. $C_a=0$, $M^2_{nucl}=0$):
\vspace{-0.2cm}
\begin{eqnarray}
\chi^2/DOF (F_2) =\frac{3258}{567},~~
\chi^2/DOF (\mbox{slope}) =\frac{1977}{7},~~
\Lambda^{f=4}_{\overline{MS}} = 300 \pm 5 ~MeV \nonumber
\end{eqnarray}
\item Only TMC are included (i.e. $C_a=0$):
\vspace{-0.2cm}
\begin{eqnarray}
\chi^2/DOF (F_2) =\frac{1145}{567},~~
\chi^2/DOF (\mbox{slope}) =\frac{464}{7},~~
\Lambda^{f=4}_{\overline{MS}} = 257 \pm 5 ~MeV \nonumber
\end{eqnarray}
\item TMC and $h_4^{free}$ (see Table) are included:
\vspace{-0.2cm}
\begin{eqnarray}
\chi^2/DOF (F_2) =\frac{369}{560},~~
\chi^2/DOF (\mbox{slope}) =\frac{15}{7},~~
\Lambda^{f=4}_{\overline{MS}} = 287 \pm 28~ MeV \nonumber
\end{eqnarray}
%The values of $h_4^{free}$ are given in the Table 1.
%
\item TMC, $C_4^{der}$ and $h_6^{free}$ (see Table) are included:
\vspace{-0.2cm}
\begin{eqnarray}
\chi^2/DOF (F_2) &=&\frac{462}{559},~~
\chi^2/DOF (\mbox{slope}) =\frac{70}{7},~~
\Lambda^{f=4}_{\overline{MS}} = 154 \pm 22 ~MeV \nonumber \\
A^{der}(H_2) &=& 0.244 \pm 0.041 ~GeV^2,~~
A^{der}(D_2) = 0.162 \pm 0.037 ~GeV^2 \nonumber
\end{eqnarray}
%The values of $h_6^{free}$ are given in the Table 1.
%
%
%\item TMC, $C_4^{ren}$ and $\overline h_6^{free}$ (see Table) are included:
%\begin{eqnarray}
%\chi^2/DOF (F_2) &=&\frac{374}{549},~~
%\chi^2/DOF (\mbox{slope}) =\frac{13}{7},~~
%\Lambda^{f=4}_{\overline{MS}} = 380 \pm 16 ~MeV \nonumber \\
%A^{ren}_4(H_2) &=& 0.160 \pm 0.016 ~GeV^2,~~
%A^{ren}_4(D_2) = 0.182 \pm 0.018 ~GeV^2 \nonumber
%\end{eqnarray}
%%The values of $\overline h_6^{free}$ are given in the Table 1.
%%
\item TMC, $C_4^{ren}$ and $C_4^{ren}$ are included:
\vspace{-0.2cm}
\begin{eqnarray}
\chi^2/DOF (F_2) &=&\frac{560}{567},~~
\chi^2/DOF (\mbox{slope}) =\frac{15}{7},~~
\Lambda^{f=4}_{\overline{MS}} = 125 \pm 17 ~MeV \nonumber \\
A^{ren}_4(H_2) &=& 0.485 \pm 0.035 ~GeV^2,~~
A^{ren}_4(D_2) = 0.457 \pm 0.032 ~GeV^2 \nonumber \\
A^{ren}_6(H_2) &=& 0.0070 \pm 0.0005 ~GeV^4,~~
A^{ren}_6(D_2) = 0.0024 \pm 0.0006 ~GeV^4 \nonumber
\end{eqnarray}
\end{itemize} 

Let us now present the main results.
%which follow from the Table and the  figures.
%Looking carefully at the results presented in the Tables 1, 2 and on the 
%figures we can make the following conclusions
\begin{itemize}
%\item our description of the $Q^2$ evolution of the asymmetry \aa1 
%has very simple form (\ref{5}) and the results are in a good agreement with 
%standard analyses \cite{GRSV,GS96,Q2E154}.
%
\item
%Both renormalon-type and non-renormalon type
The different parametrizations of HT terms lead to very good fits of SLAC,
BCDMS and NMC data \cite{data}.
Including twist-4 term correction as free parameters at each $x_i$ bin
leads to  $\Lambda^{f=4}_{\overline{MS}}   \sim 286 \pm 28~ MeV$
(or $\alpha_s(M^2_Z) \sim 0.114 \pm 0.002$).
However,
%the renormalon-type
the HT tems, which are obtained in the framework of infrared renormalon
model and also  in the form of the derivative
$ \frac{d}{dx}lnF_2^{NS}(x,Q^2)$,
lead to the lower values of QCD
parameter $\Lambda^{f=4}_{\overline{MS}}   \sim 140 \pm 21~ MeV$
(or $\alpha_s(M^2_Z) \sim 0.103 \pm 0.002$).
It is in perfect agreement with the recent results \cite{DELPHI} of
DELPHI Collaboration, where the strong reduction of the $\alpha_s(M^2_Z)$
value was observed when renormalon-type power
corrections have been included. The reduction in \cite{DELPHI} is even
stronger because the
nonperturbative part for the cross sections of $e^+e^-$ collisions
starts \cite{DoWe} with the $1/\sqrt{Q^2}$ term. Moreover,  similar effect
of the
reduction of the $\alpha_s(M^2_Z)$ value has been also observed earlier 
in the analyses of DIS data based on NNLO approximation \cite{NNLO}
of perturbative part and also on scheme-invariant
perturbation theory \cite{SIPT} resumming a part of higher order terms.
\item  
Twist-4 terms are very similar to ones obtained in \cite{VM}, when we
considered them as free parameters.
\item
Twist-6 terms are different
%essentially
for proton and deutron SF
that seems to indicate  a nonzero nuclear correlations in deutron.
\end{itemize}
\hskip -.56cm
%
%%%\begin{table}[t]
%%%\caption
{\bf Table.} The values of 
the parameters for $h_4^{free}$ ($~GeV^2$) and $h_6^{free}$
($~GeV^4$).
%and $\overline h_6^{free}$ ($~GeV^4$).
%%%\label{tab:exp}}
\vspace{0.2cm}
\begin{center}
\footnotesize
%%%\begin{tabular}{|l|c|c|c|c|c|c|}
\begin{tabular}{|p{35pt}|p{38pt}|p{38pt}|p{38pt}|p{38pt}|p{38pt}|p{38pt}|}
\hline
%{} &\raisebox{0pt}[13pt][7pt]{$\Gamma(\pi^- \pi^0)\; s^{-1}$} &
%\raisebox{0pt}[13pt][7pt]{$\Gamma(\pi^-\pi^0\gamma)\; s^{-1}$} &{}\\
$x_i$  & 0.35 & 0.45 & 0.55 & 0.65 & 0.75 & 0.85 \\
\hline
%\hline
%\multicolumn{5}{|c|}{
%%\raisebox{0pt}[12pt][6pt]{Process for Decay}} & &
%SMC proton and deutron data}\\
%%\cline{1-2}
$h_4^{free}(H_2)$ & -0.245 $\pm$ 0.025 & -0.171 $\pm$ 0.045 &
-0.029 $\pm$ 0.076 & 0.336 $\pm$ 0.133 & 0.759 $\pm$ 0.239 &
1.41 $\pm$ 0.60 \\  
\hline
$h_4^{free}(D_2)$ & -0.205 $\pm$ 0.030 & -0.064 $\pm$ 0.048 &
0.106 $\pm$ 0.078 & 0.271 $\pm$ 0.128 & 0.426 $\pm$ 0.213 &
1.44 $\pm$ 0.55 \\  
\hline
$h_6^{free}(H_2)$ & -0.266 $\pm$ 0.018 & -0.425 $\pm$ 0.054 &
-0.618 $\pm$ 0.134 & -0.448 $\pm$ 0.414 & -1.61 $\pm$ 0.99 &
-9.52 $\pm$ 4.32 \\  
\hline
$h_6^{free}(D_2)$ & -0.198 $\pm$ 0.020 & -0.139 $\pm$ 0.057 &
0.003 $\pm$ 0.137 & 0.147 $\pm$ 0.398 & -0.651 $\pm$ 0.821 &
3.72 $\pm$ 4.36 \\  
%\hline
%$\overline h_6^{free}(H_2)$ & -0.179 $\pm$ 0.037 & 0.050 $\pm$ 0.080 &
%0.249 $\pm$ 0.137 & 0.518 $\pm$ 0.308 & -1.23 $\pm$ 0.79 &
%-10.90 $\pm$ 3.86 \\  
%\hline
%$\overline h_6^{free}(D_2)$ & -0.102 $\pm$ 0.042 & 0.287 $\pm$ 0.086 &
%0.704 $\pm$ 0.141 & 0.350 $\pm$ 0.282 & -2.26 $\pm$ 0.59 &
%-5.77 $\pm$ 3.87 \\  
\hline
\end{tabular}
\end{center}
%%%\end{table}\\
%%%\end{minipage}
%%%\end{tabular}

\vspace{-0.2cm}

\section*{Acknowledgments}
\vspace{-0.2cm}

One of us (AVK) is gratefull very much to Organizing Committee of 
Workshop DIS98
for the financial support. 

\vspace{-0.3cm}

%%                 REFERENCES %

\end{document}